\documentclass[12pt]{article}
\usepackage{graphicx}
\usepackage{amsfonts}
\usepackage{latexsym}
\usepackage{amsmath}

\makeatletter
\@addtoreset{figure}{section}
\def\thefigure{\thesection.\@arabic\c@figure}
\@addtoreset{table}{section}
\def\thetable{\thesection.\@arabic\c@table}

\def\@sect#1#2#3#4#5#6[#7]#8{\ifnum #2>\c@secnumdepth
     \def\@svsec{}\else
     \refstepcounter{#1}\edef\@svsec{\csname the#1\endcsname.\hskip .75em
}\fi
     \@tempskipa #5\relax
      \ifdim \@tempskipa>\z@
        \begingroup #6\relax
          \@hangfrom{\hskip #3\relax\@svsec}{\interlinepenalty \@M #8\par}%
        \endgroup
       \csname #1mark\endcsname{#7}\addcontentsline
         {toc}{#1}{\ifnum #2>\c@secnumdepth \else
                      \protect\numberline{\csname the#1\endcsname}\fi
                    #7}\else
        \def\@svsechd{#6\hskip #3\@svsec #8\csname #1mark\endcsname
                      {#7}\addcontentsline
                           {toc}{#1}{\ifnum #2>\c@secnumdepth \else
                             \protect\numberline{\csname the#1\endcsname}\fi
                       #7}}\fi
     \@xsect{#5}}
\def\@begintheorem#1#2{\it \trivlist \item[\hskip \labelsep{\bf #1\ #2.}]}
\def\section{\@startsection {section}{1}{\z@}{-3.5ex plus -1ex minus
 -.2ex}{2.3ex plus .2ex}{\normalsize\bf}}

\begin{document}

\title{Epitaxial Frustration in Deposited Packings of Rigid Disks and Spheres}
\date{} 
\maketitle
\vspace{0.5\baselineskip}
\hspace{-.4in}
{\begin{tabular}{lr}
{\em ~~~~~~~~~~~~~Boris D. Lubachevsky} & 
{\em ~~~~~~~~~~~~~~~~~~~~~~~~~~~~~~~~Frank H. Stillinger} \\
~~~~~~~~~~~~~~Bell Laboratories & Department of Chemistry\\
~~~~~~~~~~~~~~Lucent Technologies & Princeton University\\
~~~~~~~~~~~~~~Murray Hill, NJ 07974 & Princeton, NJ 08544 \\
~~~~~~~~~~~~~~bdl@bell-labs.com & fhs@princeton.edu
\end{tabular}

\setlength{\baselineskip}{0.995\baselineskip}

\begin{center}

{\bf ABSTRACT}
\end{center}

We use numerical simulation to investigate and analyze 
the way that rigid disks and spheres arrange 
themselves when compressed next to incommensurate substrates.  
For disks, a movable set is pressed into 
a jammed state against an ordered fixed line of larger disks, 
where the diameter ratio of movable to fixed 
disks is 0.8.  
The corresponding diameter ratio for the sphere simulations is 
0.7, where the fixed substrate 
has the structure of a (001) plane 
of a face-centered cubic array.  
Results obtained for both disks and 
spheres exhibit various forms of 
density-reducing packing frustration next to the incommensurate 
substrate, 
including some cases displaying disorder that extends far from the substrate.  
The disk system 
calculations strongly suggest that the most efficient (highest density) 
packings involve configurations that 
are periodic in the lateral direction parallel to the substrate, 
with substantial geometric disruption only 
occurring near the substrate.  
Some evidence has also emerged suggesting that for the sphere systems 
a corresponding structure doubly periodic 
in the lateral directions 
would yield the highest packing density; 
however all of the sphere simulations completed thus far 
produced some residual ``bulk'' disorder not 
obviously resulting from substrate mismatch.  
In view of the fact that the cases studied here represent only 
a small subset of all that eventually deserve attention, 
we end with discussion of the directions in which
first extensions of the present simulations might profitably be pursued.

\setlength{\baselineskip}{1.2\baselineskip}
\section{Introduction}
\hspace*{\parindent}
A diverse group of physical processes, 
both in nature and in human technological activities, 
involve
deposition of various substances on solid substrates.  
Examples of the former include accretion of low
molecular weight materials on interstellar grains at low temperature 
\cite{Greenberg}, 
and on high-altitude air-borne 
particles in the terrestrial atmosphere 
\cite{Kulmala}.  
The latter category covers the now-indispensable industrial 
practices that produce miniaturized integrated circuitry 
for computer technology 
\cite{Semiconductor}.  
But whether driven 
by pure scientific curiosity or by economic and societal needs, 
understanding the subtleties of deposition phenomena in surface physics 
generally remains a challenging subject.  
The present paper is intended to 
contribute in at least a modest way to deepening that understanding.

The specific objective behind our study is 
clarification of geometric frustration effects in deposition processes 
that can arise from size discrepancy 
between particles of the solid substrate, 
assumed here to be crystalline, 
and particles of the deposited layer.  
In order to attain at least a simplified view of such 
frustration effects, 
we have chosen to examine the idealized 
rigid disk (two dimensional) 
and rigid sphere (three dimensional) 
models for the operative interparticle interactions.  
The present investigation is an extension of our previous publication 
that investigated the statistical geometry of disk and sphere packings, 
but where the substrate particle sizes were identical 
to those of the added layer \cite{SL}.  
Now we relax that restriction.  
The simulation procedure again 
involves continuous compression of the depositing material, 
initially in a random fluid state, 
toward the substrate until jamming occurs.  
The incommensurate substrate used in the present study 
is represented as a fixed crystal layer 
containing no defects (screw dislocations, missing particles, etc.).  
However, the size mismatch between this layer and the particles
forced into contact with it 
creates stacking disorder in the deposited layer, 
the types and amounts of which are a focus of the present investigation.

Of course representing real atomic, 
ionic, 
or molecular interactions as those operating 
between rigid disks or spheres 
is a drastic simplification.  
Nevertheless, it is a simplification with a long and worthy history.  
With respect to jammed arrangements of disks and spheres, 
one can cite a broad array of published papers 
that involve 
experimental 
\cite{Bernal, SK, PNW}, 
simulational 
\cite{SL, VB, JT, TC, LS, TTD}, 
and analytical 
\cite{Berryman, Connelly88, Connelly91, TS, TDS}
techniques.  
In many of the past studies, 
care has been exercised to eliminate boundary effects; 
by contrast here we specifically focus on a special class of those effects.

The following 
Section 2 
presents details of our numerical simulation procedure.  
These details include choices about system sizes, 
particle size ratios and epitaxial surface geometries, 
compression rates, and jamming criteria.  
Section 3 
collects numerical results for the two-dimensional 
disk calculations emerging from that procedure.  
Section 4
does the same for the three-dimensional sphere cases.  
The final 
Section 5 
includes several conclusions based on the results; 
it also offers discussion and speculation about 
several features of this class of epitaxial packing phenomena that, 
in our opinion, 
deserve future research attention.

\section{Simulation Procedure}
\hspace*{\parindent}
For both the two and three dimensional applications, 
the system geometry to be used must present the incommensurate substrate 
to the movable adsorbing particles, 
and must provide a means for tightly compressing 
the collection of those movable particles against that substrate.  
In order to satisfy these requirements, 
a rectangular (two dimensions) 
or rectangular solid (three dimensions) 
cell is utilized, 
one face of which is the structurally fixed incommensurate substrate.  
The opposite face also consists of 
a structurally fixed array 
of disks or spheres 
that are in contact with one another, 
but it is composed of particles 
equal in size to those of the movable set; 
this choice avoids further incommensurability 
and roughly simulates the effect 
of an infinite system depth in the direction 
normal to the substrate surface.  
The distance between these two faces is variable, 
and thus permits implementation of the required compaction process.  
The remaining two faces (for disks) 
or four faces (for spheres) 
retain fixed separation, 
and are treated as periodic boundary conditions. 
\begin{figure}
\includegraphics*[width=6.2in]{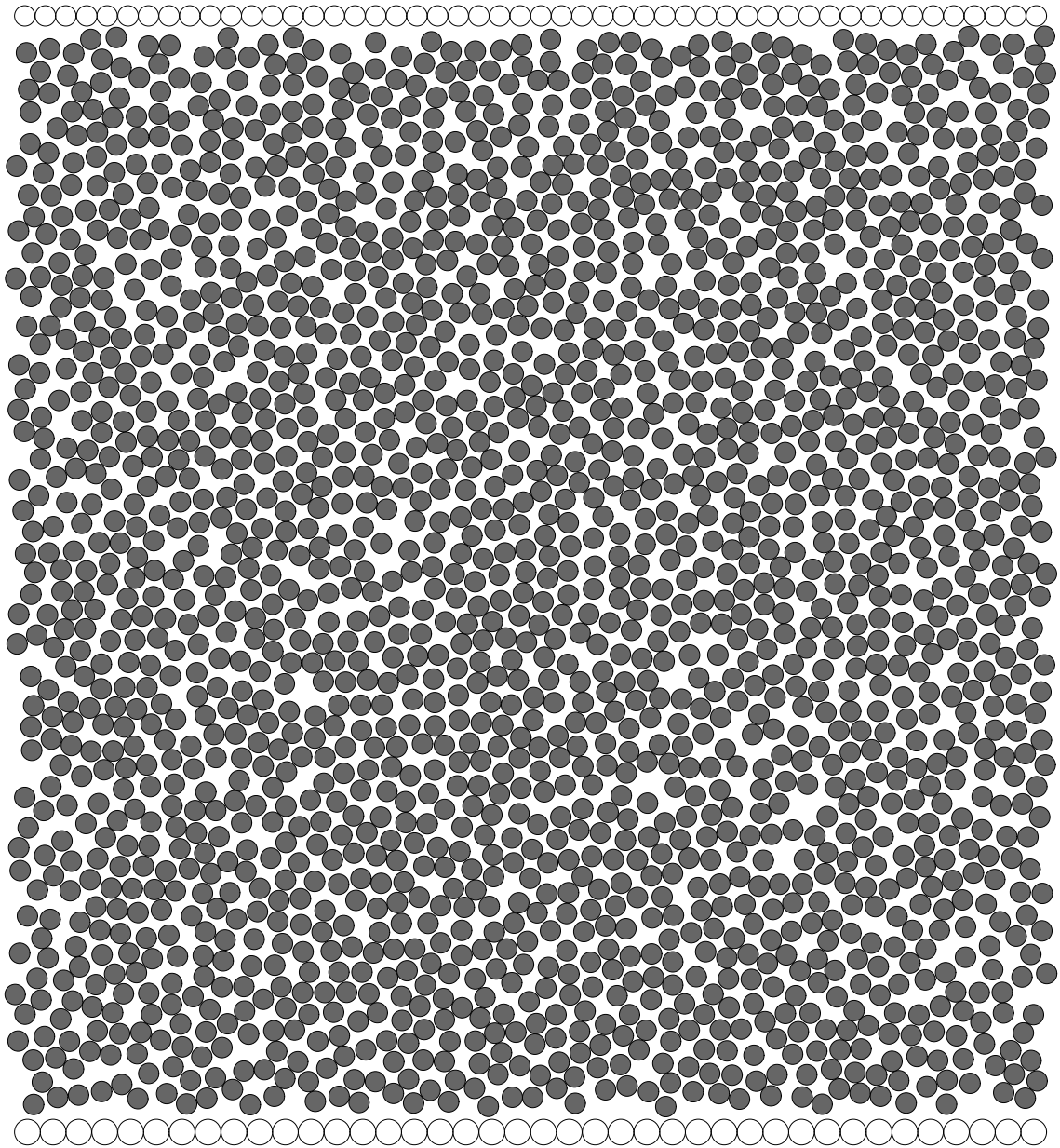}
\caption{Initial state of a disk simulation, case D, 
comprising 40 large disks as a fixed incommensurate bottom array, 
50 smaller disks fixed at the top, 
and 2000 movable disks of the smaller size.  
The diameter ratio is 0.8.  
Periodic boundary conditions apply 
in the lateral direction parallel to the fixed layers
}
\label{initial}
\end{figure}

For purposes of illustration, 
Figure ~\ref{initial} 
provides a view of the initial stage of a disk compression.  
This example involves 
a linear arrangement of 40 fixed 
contacting disks acting as the incommensurate bottom 
boundary. 
The diameter of these fixed disks 
serves as the unit length for the calculations.  
The remaining disks have diameters equal only to 0.8 times this unit length.  
These 2050 smaller disks include 50 that 
form the rigid top layer, 
and 2000 that are individually free to move.  
The lateral distance $L_x$ between the left and right periodic boundaries 
remains equal to 40 length units 
(i.e. 50 smaller-disk diameters) during
the subsequent numerical simulation.  
If in fact the fixed bottom layer also comprised 50 smaller disks,
vertical compression 
could in principle produce 
a perfect triangular array consisting of a total of 42
horizontal rows, 
on account of the choices of movable disk number 
and lateral boundary separation.  
Among 
all possible arrangements of non-overlapping disks, 
the perfect triangular array 
(infinitely extended) 
attains the largest possible covering fraction:
\begin{equation}
\label{hexden}
\xi_{\max}(D=2) ~=~ \pi / (2 \sqrt{3}) ~=~ 0.906899682...
\end{equation}

In order 
to prepare an initial configuration 
such as that shown in Figure ~\ref{initial}, 
we followed a procedure closely related 
to that employed in our earlier study 
\cite{SL}.  
This preparation stage first involves growing
2000 disks at a fixed rate 
from randomly placed points, 
while allowing for elastic collisions to avoid 
overlap of pairs of expanding disks.  
Disk growth is terminated when the target diameter 0.8 is attained.
A detailed description of this growth process 
is available in reference 
\cite{LS}.  
In the present case, 
the growing disks are confined 
to a fixed-size rectangle $L_x \times L_y$, 
subject to periodic boundary conditions 
in both directions, 
with $L_x = 40$ , and $L_y$   
initially selected so that the 2000 fully-grown disks 
within the initial 
area  $L_x L_y$
have a covering fraction equal 
to  $(2/3)~\xi_{\max}~(D=2)$.
This requires 
choosing  $L_y(initial)~=~(2 \sqrt{3})^3~=~41.5692...$
at this preparation stage.

Having generated this moderate-density configuration 
for the 2000 movable disks, 
it is then necessary 
to bound them at bottom and top respectively 
with the structurally-fixed layers of 
40 unit-sized substrate 
disks, 
and 50 smaller-sized (diameter 0.8) disks.  
This is done simply by placing those layers directly below 
(by 0.9 units) 
and directly above 
(by 0.8 units) 
the lower and upper boundaries of the preceding 
rectangular cell, 
thus just avoiding any overlaps.  
Subsequent to this addition,  
$L_y$ is measured from the 
centers of those lower and upper disk layers, 
consequently increasing this dimension by 1.7 units. 
The resulting distance between the centers of 
the lower and upper sets of bounding disks at the beginning of 
the compression phase, as illustrated in Figure ~\ref{initial}, is thus
\begin{equation}
\label{Lyinit}
L_y(t=0)~=~43.2692...
\end{equation}

The 2000 movable disks 
(now with fixed diameters) 
between the structurally rigid layers inherited 
velocities from the initial setup stage, 
which were then normalized so as to have mean speed equal to
unity.  
Subsequently, 
particle motions caused 
elastic collisions between pairs of movable disks, and 
between movable disks and the top and bottom layers.  
Simultaneously, compression toward a jamming
limit was effected by moving the upper and lower layers 
symmetrically toward one another 
so that the vertical system dimension 
declines at the following constant rate: 
\begin{equation}
\label{ypressrate}
dL_y/dt~=~-0.001
\end{equation}
On the basis of prior experience 
\cite{SL}, 
this rate is 
sufficiently small 
that any final-stage disorder is expected 
to result primarily 
from incommensurabilty at the substrate, 
not from random jamming processes arising 
in the interior of the disk system.  
As a purely mathematical observation, 
in terms of the time-like 
progress 
variable $t \ge 0$, 
compression would cause the collision rate to increase, 
and formally to diverge as the jammed state is approached 
\cite{SL, LS}.  
However, 
practical numerical considerations require 
(a) scaling 
particle velocities downward at late stages of the compression, 
and 
(b) terminating any run after a finite, 
but large number of collisions, 
chosen so that 
in fact the particle system is geometrically very 
close to its ideally jammed state.  
Our simulations consisted of series of dynamical ``sessions'', 
each containing a 
number of collisions 
equal to an average per particle of 
$10^3$; 
if 10 successive sessions produced reductions 
in the covering fraction by less 
than $10^{-10}$, 
the compression to jamming was numerically regarded as 
complete.

Five distinct examples of the disk-system type 
just described have been processed.  
They differed only in the random number seeds 
used to initiate the runs.  
In addition to these, 
five more disk cases were examined 
in which only 1990 small movable disks (diameter 0.8), 
rather than 2000, were present.  
The motivation 
for reducing the movable set by 10 
was that the lower bounding substrate presented 
10 fewer ``pockets'' than those of the upper bounding layer, 
and it was of interest to see what effects, 
if any, 
this modification would produce in the final jammed structures.  
In order 
to have the same initial covering fraction 
with this reduced number of movable disks, 
the initial cell 
dimension $L_y$
was correspondingly slightly reduced.

The initial conditions 
employed for the three-dimensional 
sphere simulations were analogous to those 
for the disk examples.  
The structurally rigid substrate surface, 
composed of unit-diameter spheres, 
consisted of a (001) face 
of a face-centered cubic close-packed array.  
This amounts to a set of mutually contacting spheres 
in a simple square planar arrangement.  
The movable spheres, 
as well as those 
comprising the upper bounding layer 
that is also a simple square planar array, 
possessed a smaller 
diameter equal to 0.7 length units.  
The lower (substrate) 
square array contained 441 unit spheres in a 21$\times$21 pattern.  
The upper layer 
square array contained 900 smaller spheres in a 30$\times$30 pattern.

Two choices were made 
for the total numbers of mobile smaller spheres, 
similarly to the disk cases, 
but necessarily much larger in magnitude.  
The specific values selected were 18,000 and 17,541.  
The former choice in principle could form a perfect 
close-packed crystal between the bounding layers upon compression, 
if the lower substrate layer were also composed of 900 small spheres.  
The latter choice 
involves 459 $(=~900-441)$  
fewer small spheres, 
reflecting the smaller number of square ``pockets'' in the substrate layer.  
On account of these much larger particle numbers 
compared to the disk cases, 
it was practical to simulate 
only single examples 
of each of these two kinds of sphere systems.  
The initial preparation stage 
grew the required number of movable spheres 
from randomly generated points 
to the final common diameter 0.7, 
in a rectangular-solid region with vertical 
length $L_z$
chosen so that the 
resulting non-overlapping spheres were present at covering 
fraction  $(2/3)\xi_{\max}(D=3)$.
Here, the maximum covering fraction for spheres, 
attainable in an infinitely extended face-centered cubic packing, 
is equal to:
\begin{equation}
\label{fccden}
\xi_{\max}(D=3) ~=~ \pi / (3 \sqrt{2}) ~=~ 0.7404804897...
\end{equation}

Following this growth stage, 
the lower and upper bounding crystal surfaces 
were put in place at bottom 
and top of the initial rectangular-solid cell, 
thus increasing the vertical dimension by 1.55 units. 

Immediately after this sphere-system preparation process, 
the velocities of the movable particles were 
renormalized to have mean magnitude unity.  
The subsequent compression phase of the simulation 
involved the following rate of height reduction:
\begin{equation}
\label{zpressrate}
dL_z/dt~=~-0.001
\end{equation}
As in the case of disks,
this compression was presumed to be slow enough 
to permit kinetic processes prior to jamming 
to anneal out most disorder arising from 
bulk-system packing irregularities.  
Substantially the same kind of termination criterion 
as used for the disk simulations 
was invoked to decide numerically 
when the sphere compression stage was ``complete''.  

One of the fundamental properties 
to be determined for each simulated jamming case is the final 
covering fraction.  
Such results need to be interpreted in relation to the maximum 
values  $\xi_{\max}(D=2)$
in Eq. \eqref{hexden}, 
and  $\xi_{\max}(D=3)$
in Eq. \eqref{fccden}, 
which refer to optimal packings of infinite extent.  
By contrast, 
the present study involves finite systems 
and emphasizes the effects exerted by boundaries 
in those finite systems.  
Several different conventions are possible 
when defining the covering 
fraction   
for a finite 
disk or sphere sample 
with boundaries of the type specified above.  
The most straightforward convention would report the value of:

\begin{equation}
\label{sfdenatt}
\begin{array}{rl}
{\xi~(t)~=~A_{total}}~/~L_x L_y (t)~~~~ & {\mbox{(D=2, disks)}} \\
~~~~~~~= V_{total}~/~L_x L_y L_z (t) & {\mbox{(D=3, spheres)}}
\end{array}
\end{equation}
for the covering fraction at compression 
time $t\ge0$.  
Here,   
$A_{total}$ and $V_{total}$
respectively denote the total 
disk area, 
and total sphere volume, 
occupied by all particles or portions thereof, 
lying within the 
rectangular or rectangular solid system region.  
This includes the entire area or volume covered by 
movable particles, 
but only semicircles or hemispheres for those particles 
comprising the structurally fixed lower substrate and upper boundary layers.  
However, for some purposes it is desirable to eliminate 
the lower and upper bounding-layer particles from consideration, 
thus focusing attention entirely on the movable set. 
A rational way to do so would be to 
replace   
$A_{total}$ and $V_{total}$
with  
$A_{mobile}$ and $V_{mobile}$,
and 
to diminish the cell area or volume 
by an amount attributable to the boundary layer 
``half-particles'' in their close-packed configurations.  
Consequently, 
this alternative choice for the covering fraction 
definition would have the form:

\begin{equation}
\label{mddenatt}
\begin{array}{rl}
{\hat{\xi}(t)~=~~A_{mobile}}~/~[~L_x L_y (t)~-\Delta A~] & 
{\mbox{(D=2, disks)}} \\
~~~~~~~~= V_{mobile}~/~[~L_x L_y L_z (t)~-\Delta V~] & {\mbox{(D=3, spheres)}}
\end{array}
\end{equation} 
Here  $\Delta A$ and $\Delta V$   
are the areas and volumes 
attributable to the boundary layer half-particles in their 
respective close-packed arrangements.  
Numerical values utilizing both definitions will be quoted below 
in discussing our results. 

In addition to the covering fractions 
as an overall measure of the jammed packing geometry, 
several 
other more detailed measures have also been evaluated.  
These include spatial distributions of particle 
collision rates in the final stage of the jamming process, 
the stratification of the density of particle centers
in the direction normal to the substrate surface, 
and the distribution of numbers of contacts 
experienced by particles in the final state.  
More global features of the final configurations are best conveyed by
pictorial presentations of the jammed systems, 
and representative examples are presented in the following two Sections.

\section{Numerical Results, Disks}
\hspace*{\parindent}
Table~\ref{tab1} 
reports the final covering 
fractions 
$\xi$  and   
$\hat{\xi}$
for the ten runs of disk simulations.  
These runs, 
designated "A" through "J", 
have been classified by $N$, 
the number of mobile disks.  
The last column in 
Table~\ref{tab1}
provides a rough descriptive category 
for the type of geometric frustration exhibited 
by the final-state 
jammed configuration; 
details of these various frustration patterns appear below.

Figure ~\ref{1cased} 
shows the final state after compression 
for case D 
whose initial configuration appears in 
Figure ~\ref{initial} 
A pair of simple graphs 
at the right edge of this picture 
present additional information about the configuration, 
as a function of the vertical coordinate, 
i.e. perpendicular to the structurally fixed bounding layers.  
The multi-peaked curve 
just to the right of the picture provides 
a view of the vertical distribution 
of mobile particle centers.  
The individual bins 
used to accumulate data 
for this graph 
have width equal to 
one-tenth the expected period 
for an undistorted crystalline arrangement of the smaller disks, 
namely 0.06928... units.  
The magnitude of the graph's peaks can be deduced
with the help of the
ticks at the graph's top edge.
The $x$-coordinate of the leftmost tick corresponds
to 0, i.e., to the value ``no particles in the bin''.
The ticks follow with horizontal spacing of 10\% of 50;
note that 50 is the maximum number of particles 
whose centers can be found at a particular height.
Thus, the second tick corresponds to 10\%, i.e., to 5 particles in a bin,
the third to 20\%, i.e., to 10 particles, and so on.
It is clear that, 
for this one example at least, 
the main disruption of the vertical-direction 
density distribution 
is strongly localized 
in the first few layers 
next to the larger-disk bottom layer.  

The farther-right curve exhibits 
the vertical variation of contact and near-contact numbers 
for the movable particles, 
using vertical bin width equal to 
twice the ideal crystal spacing, 
and connecting the results with 
line segments for ease of visualization.  
The magnitude of this graph can be deduced
with the help of a different set of ticks
at the graph's top edge, where again the leftmost tick
corresponds to 0, i.e., to the value ``no contact'',
and the horizontal spacing between consecutive ticks 
corresponds to one contact.
Thus, the second tick corresponds to one contact,
the third to two contacts, and so on.
The criterion for contact or near-contact involved identifying 
neighbors with their surfaces 
no further separated from the particle of interest 
than approximately 
$10^{-3}$ 
disk diameters.  
As expected, 
the values shown 
tend to be depressed below the 
maximum possible (6) 
in layers with substantial geometric disruption.

The 2000 disks 
that were mobile during the compression phase of case D 
have been distinguished in 
Figure ~\ref{1cased} 
by shading according to their relative collision rates 
during the final stages of approach to jamming. 
The lighter-shaded disks 
have experienced higher collision rates 
than the darker-shaded disks.
The latter category includes 
a few ``rattlers'' 
(trapped but unjammed disks) 
with very low collision rates.
For consistency, 
the mutually-contacting fixed particles 
comprised in the top and bottom layers 
are unshaded as though they had infinite collision rates. 
Examination of 
Figure ~\ref{1cased} 
shows that disks near the 
bottom boundary have exhibited lower collision rates 
than those at higher elevations.  
In view of the fact that geometric frustration 
is concentrated near that bottom layer, 
the presence of this collision rate 
distinction should not be surprising.  
Even excluding the rattlers 
that are concentrated in the lower layers, 
the shading distinctions displayed in 
Figure ~\ref{1cased} 
for movable disks 
span a spread of average collision rates of 
approximately two orders of magnitude.
\begin{table}
\begin{center}
\fbox{
\begin{tabular}{ c| c| c| c| l}
  N    & Case  &  $\xi$     & $\hat{\xi}$    & Category    \\  \hline\hline
  2000 & A     &  0.903169  & 0.903065  & localized   \\  \hline
  2000 & B     &  0.903160  & 0.903056  & localized   \\  \hline
  2000 & C     &  0.902808  & 0.902694  & delocalized \\ \hline
  2000 & D     &  0.903189  & 0.903085  & localized   \\ \hline
  2000 & E     &  0.894369  & 0.894021  & delocalized \\ \hline
  1990 & F     &  0.898742  & 0.898514  & delocalized \\ \hline
  1990 & G     &  0.898783  & 0.898556  & delocalized \\ \hline
  1990 & H     &  0.898790  & 0.898563  & delocalized \\ \hline
  1990 & I     &  0.898648  & 0.898417  & delocalized \\ \hline
  1990 & J     &  0.898861  & 0.898636  & delocalized 
\end{tabular}
}
\caption{
Disk jammed packing results for epitaxially-frustrated systems.  
The number of mobile particles is $N$, their radius is 0.8 times 
those of the bottom bounding layer, and the final covering fractions 
are   Eq.~\eqref{sfdenatt} and   Eq.~\eqref{mddenatt}. 
The geometric character of the epitaxial frustration 
is briefly identified in the last column
}
\label{tab1}
\end{center}
\end{table}

\begin{figure}
\includegraphics*[width=6.2in]{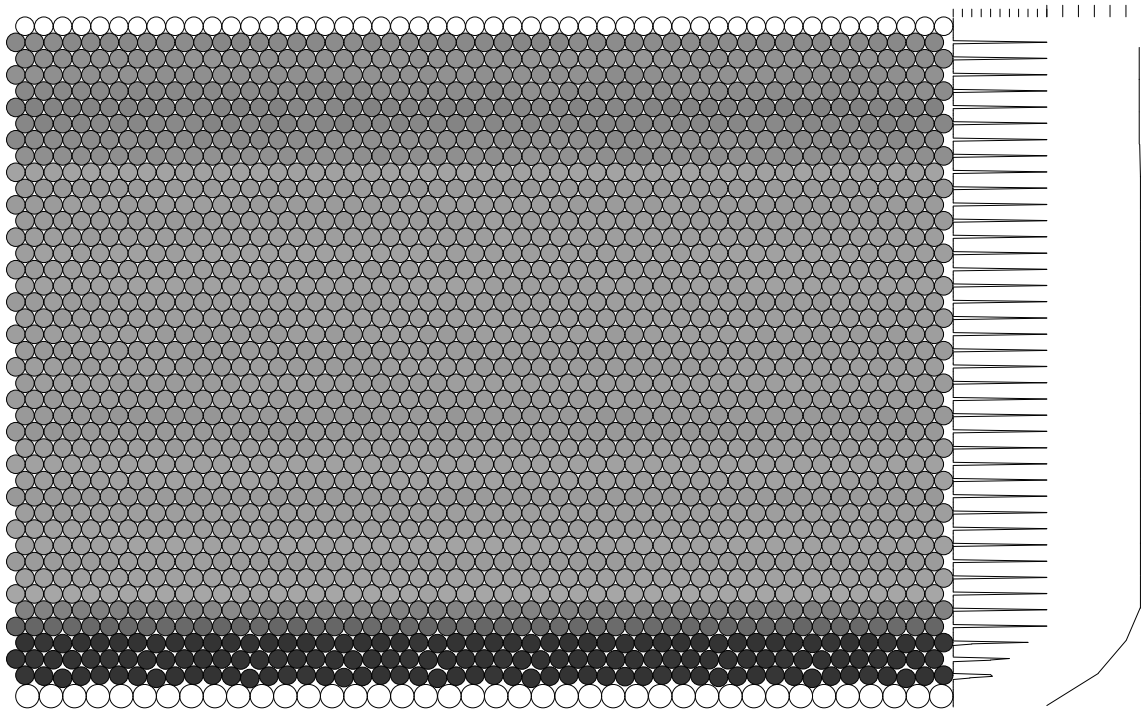}
\caption{
Final jammed configuration for the compression run 
whose initial configuration appears in Figure ~\ref{initial}. 
This case D involves 2000 mobile disks.  
A bar graph just to the right shows the vertical stratification of 
disk centers.  
The farther-right curve indicates values of mean numbers of contacts 
and near-contacts for particles in a vertical sequence of bins.  
Shading of disks indicates relative collision rate 
upon approach to the jamming limit, as explained in the text
}
\label{1cased}
\end{figure}

\begin{figure}
\includegraphics*[width=6.2in]{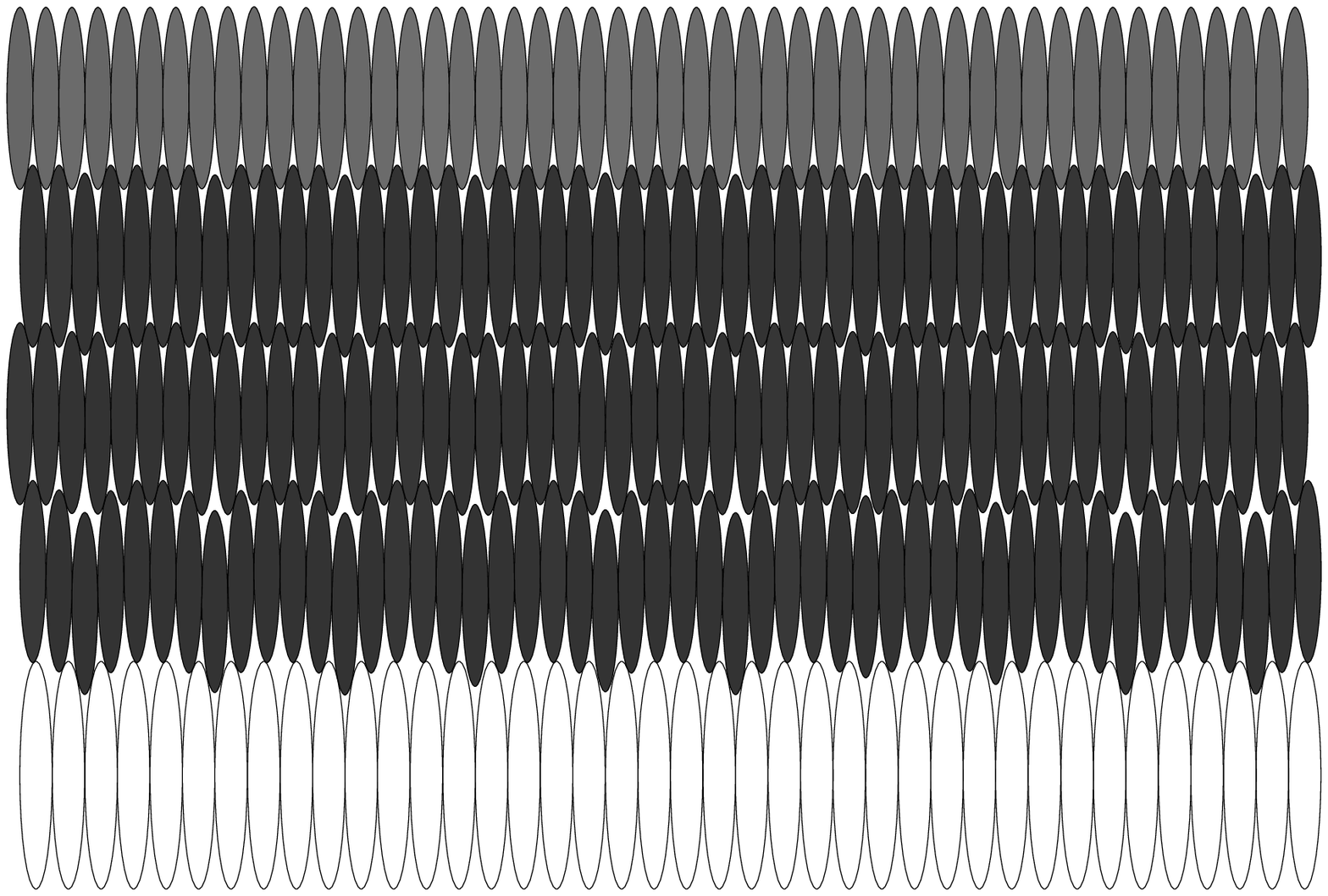}
\caption{
Vertically stretched version (by factor 7) of the bottom 
few layers from the jammed disk configuration 
shown in Figure ~\ref{1cased} for case D.  
This view emphasizes the geometric disruption 
resulting from size mismatch
}
\label{7cased}
\end{figure}
In order 
to clarify the detailed character 
of the frustrating size mismatch, 
Figure \ref{7cased} 
shows a vertically 
expanded view of the disrupted first few layers, 
from case D in 
Figure~\ref{1cased} 
just above the large-disk lower bounding layer.  
The affine transformation used to create this view 
has stretched the vertical coordinate by 
a factor 7.  
The most important feature emphasized in this manner 
is that the epitaxial disruption tends to produce 
a horizontal periodicity of four distance units, 
in which members of the lowest small-disk layer 
alternate between fitting well down into the large-disk pockets, 
and being forced upward in varying 
degrees toward sitting atop large disks.  
This near-periodicity 
(length 4 is equivalent to 5 small-disk diameters) 
has the size ratio as its obvious source. 
Figure~\ref{7cased}
makes it clear that the average amplitude 
of vertical displacement for this periodic pattern 
is relatively large for this first small-disk layer, 
but decreases rapidly 
with succeeding layers progressing upward.  
This feature is consistent with the collision rate 
differences noted above. 
Layers subject to a spatially oscillatory vertical deformation 
have a larger contour length 
than those that are not deformed 
(and are thus horizontally straight). 
As a result, 
their disks 
lie farther apart on average, 
so collisions with lateral neighbors 
in these deformed layers 
would be less frequent 
than in a perfectly aligned layer. 

Table~\ref{tab1} 
shows that the case D configuration illustrated in 
Figures~\ref{1cased} and \ref{7cased} 
nominally possesses the 
largest covering fraction 
of all five cases considered with 2000 mobile disks.  
Two other cases, 
A and B 
have slightly smaller, 
but similarly large covering fractions.  
Not surprisingly, 
examination of their final configurations 
shows that they also contain 
very nearly the same pattern 
of epitaxial frustration, mostly 
localized to the bottom few layers, 
in a horizontally nearly periodic fashion.  
The last column of 
Table~\ref{tab1} 
acknowledges this localization property for these three cases. 
The binned distributions of vertical-direction densities, 
and of contact and near-contact numbers, 
and the shading patterns for collision rates, 
are visually very similar to those shown in 
Figure~\ref{1cased},
and so need not be separately presented pictorially.  
In fact, 
it is reasonable to suppose 
that A,B, and D 
are all converging toward exactly the same 
final jammed structure, 
but have not quite attained it on account of 
finite computing-time constraints.

The consistently 
high densities observed for cases A, B, and D, 
each exhibiting localized near-periodicity of disruption, 
suggests that the highest attainable density 
(for the given particle number, particle diameter, 
and $L_x$) 
might involve strict periodicity 
in the horizontal direction with period 4.  
Note 
that the setup of our simulations 
adapts easily 
to enforcement of such horizontal periodicity.  
We simply 
reduce the configuration sizes tenfold: 
The number of movable particles becomes 200, 
the horizontal dimension 
becomes equal to the horizontal period 4, 
and the numbers of particles in the bottom and top 
rigid layers 
become 4 and 5 respectively.  
This strategy is equivalent to constraining 
a 2000-particle simulation 
to a periodic subspace of its full configuration space, 
with a consequent substantial 
computational simplification and speedup.

In general, 
a reduction in the configuration space 
would decrease or leave unchanged the maximum 
attainable density.  
However, 
the evidence presented by cases A-E 
indicates that for the unconstrained 
2000-particle system 
a large number of sub-optimal 
jammed configurations exists, 
making it unsure that 
random search 
would automatically discover 
an optimal perfectly periodic structure.  
In fact, 
we have run a few reduced-size 200-particle simulations as described, 
and the final configurations were always the same, 
with patterns 
just as suspected from examination of cases A, B, and D.  
Besides being strictly
periodic with a 4-unit period, 
this suspected optimum packing 
apparently also possesses a mirror 
symmetry with respect to vertical axes. 
These vertical mirror-symmetry axes have a spacing of 2 units 
(i.e. one-half period) 
in the horizontal ($x$) direction.  
The values for the two covering fraction definitions, 
for these ideal period-4 jammed configurations, 
were found to be  
$\xi(N=200)~=~$0.903250  and 
$\hat{\xi}(N=200)~=~$0.903148.  
These magnitudes slightly exceed the 
Table~\ref{tab1} 
entries for A, B, and D as 
initially suspected.

The remaining 
two 2000-mobile-disk jammed configurations listed in 
Table~\ref{tab1} 
(cases C and E) have 
significantly lower covering fractions.  
Figure ~\ref{1casee} 
presents the final configuration obtained for the latter
member of this pair.  
This view exhibits two linear mismatch seams in an inverted ``V'' arrangement, 
showing poor alignment 
with neighboring diagonal lines of particles 
(and thus showing gaps), 
extending 
more than halfway from bottom to top.  
The vertex of this ``V'' displays two relatively large pentagonal 
holes.  
The same collision rate shading rule has been used in 
Figure ~\ref{1casee} 
as in the prior 
Figure ~\ref{1cased}.
The vertical 
density distribution, 
and the contact and near-contact distribution, 
have again been provided at the right
margin.  
In this case the former shows peak splitting 
due to the presence of the inverted ``V'', 
and the 
latter a more complex pattern than before.  
Figure ~\ref{7cased} 
shows the bottom few layers for this configuration E, 
with factor-7 vertical stretching to emphasize 
the geometric frustration details.  
The two cases C and E 
have been designated as ``delocalized'' in the last column of 
Table~\ref{tab1}, 
owing to the fact that for both, 
disruption of the perfect triangular packing of disks 
has penetrated far above the bottom layer.  
This longer-range disruption explains 
the lower densities obtained for this pair of cases.

Table~\ref{tab1}, 
last column, 
indicates that each of the five cases F-J with 1990 disks 
is ``delocalized''.  
All of 
these have relatively low covering 
fractions $\xi$ and $\hat{\xi}$.  
Figure ~\ref{1casej} 
presents the final configuration for case J.
Figure ~\ref{7casej} 
shows the vertically stretched version 
for its bottom few layers.  
The collision-rate shading 
pattern and the plots for vertical-density, 
as well as contact and near-contact numbers, 
in the first of these 
two Figures make it very evident 
that geometric frustration pervades virtually the entire vertical 
dimension of the jammed final state.  
However, 
the specific disruption patterns exhibited 
by these cases F to J are diverse, 
generally not closely resembling one another.  
This observation suggests that the 1990-disk system 
has a significantly larger number of distinct accessible jammed states 
than does the 2000-disk system, 
the majority of which incorporate delocalized disruption.  
To illustrate this diversity, 
Figures ~\ref{1caseh} and ~\ref{7caseh}
present the final configuration 
and its stretched lower portion for case H to be compared 
with J, 
Figures ~\ref{1casej} and ~\ref{7casej}.
Evidently, 
removal of 10 mobile disks from the initially-examined 2000 can produce 
profound changes in the final geometric patterns.
\begin{figure}
\includegraphics*[width=6.2in]{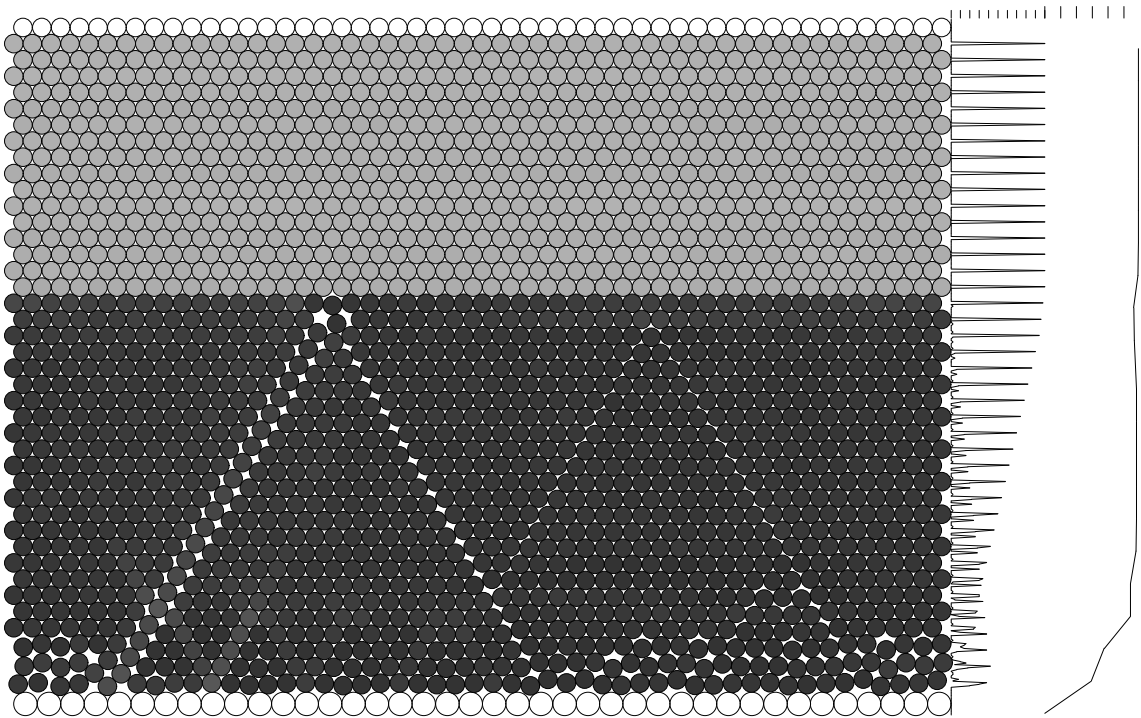}
\caption{
Final jammed configuration for the 2000-mobile-disk compression run E
}
\label{1casee}
\end{figure}

\begin{figure}
\includegraphics*[width=6.2in]{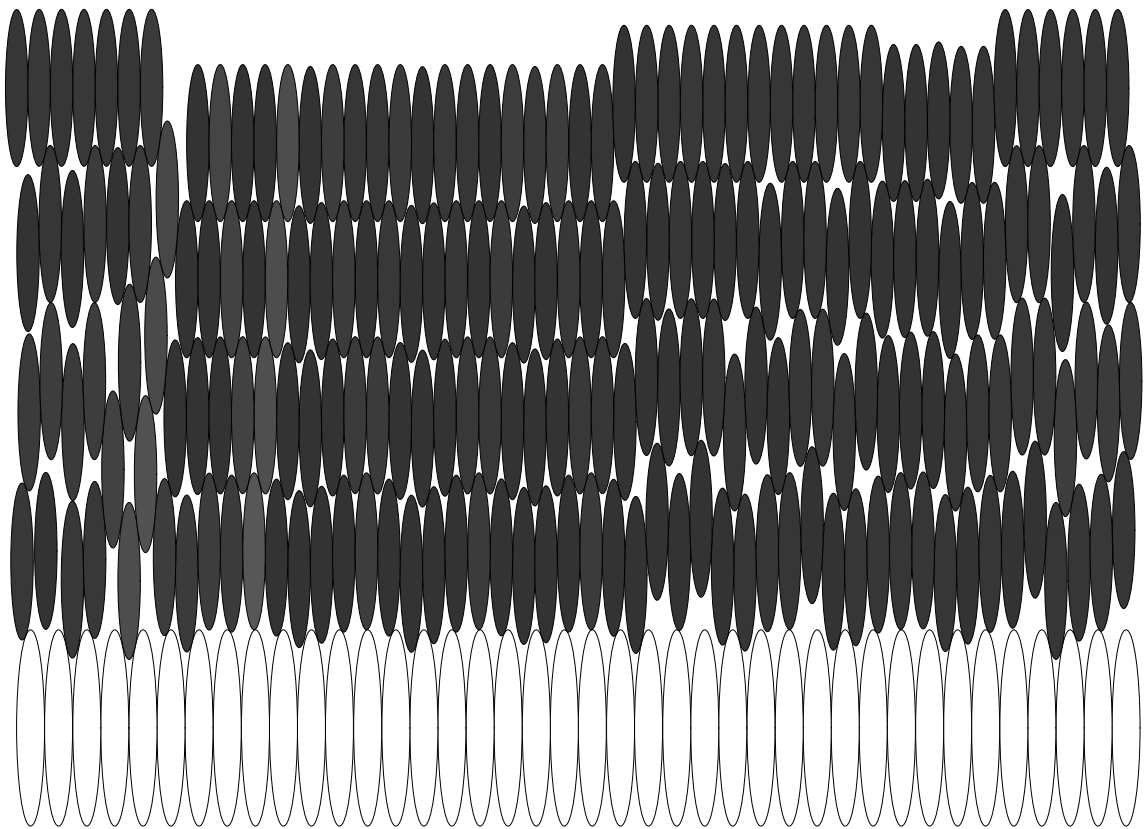}
\caption{
Vertically stretched version of the bottom few layers 
from the jammed disk configuration shown in Figure ~\ref{1casee} for case E
}
\label{7casee}
\end{figure}
\begin{figure}
\includegraphics*[width=6.2in]{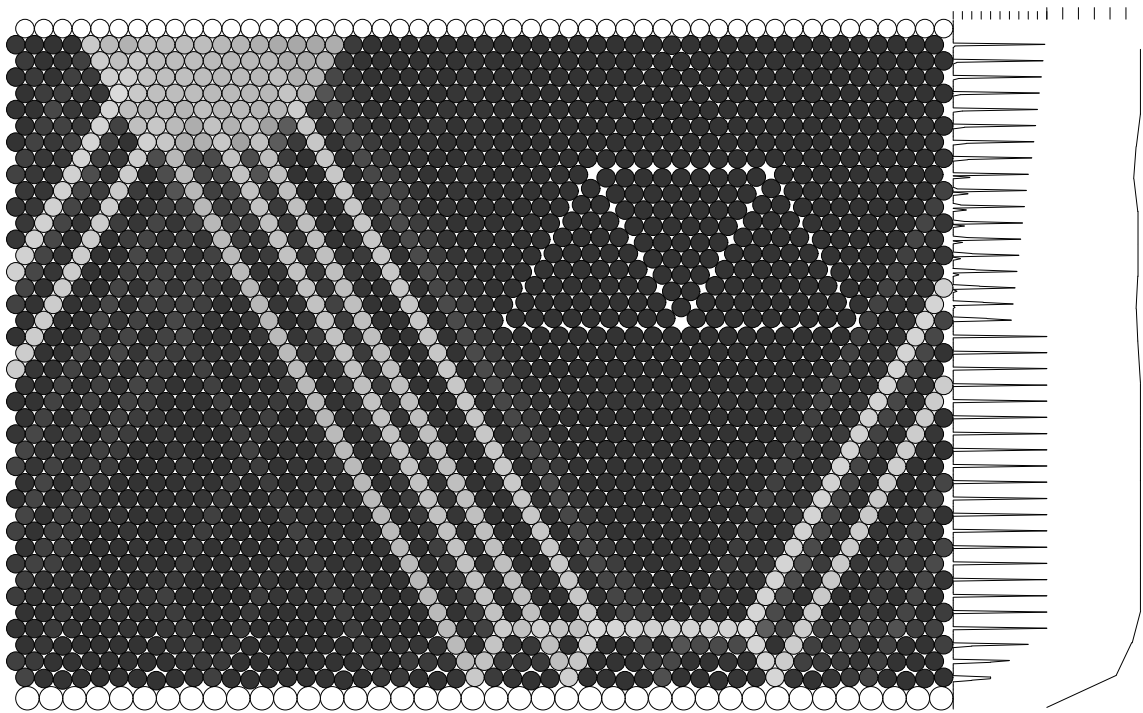}
\caption{
Final jammed configuration for case J, 
exhibiting geometric frustration intruding across 
the entire vertical dimension of this 1990-mobile-disk system
}
\label{1casej}
\end{figure}

\begin{figure}
\includegraphics*[width=6.2in]{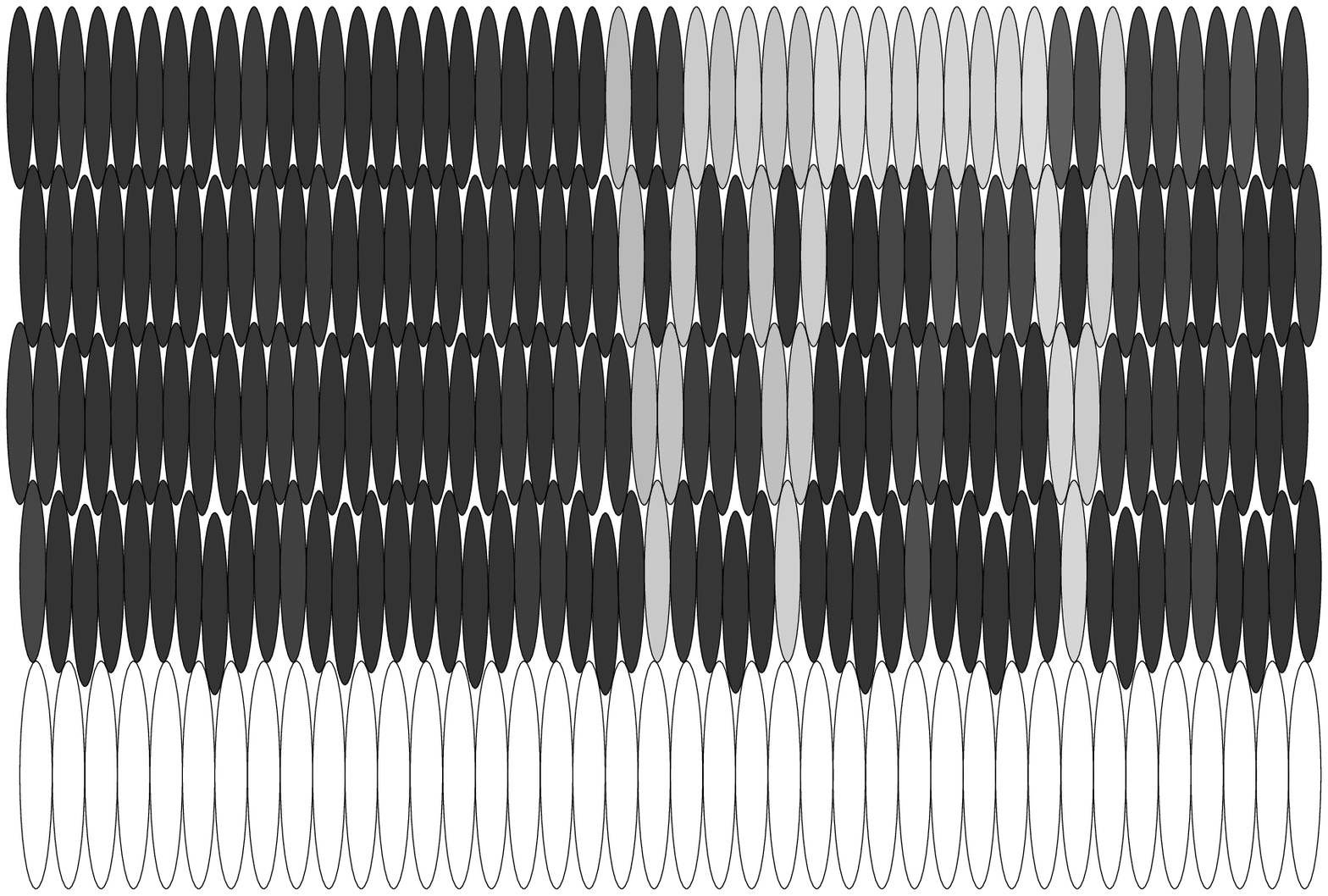}
\caption{
Vertically stretched view of the bottom layers 
in the jammed configuration for case J, 
shown in the previous Figure ~\ref{1casej}
}
\label{7casej}
\end{figure}
\begin{figure}
\includegraphics*[width=6.2in]{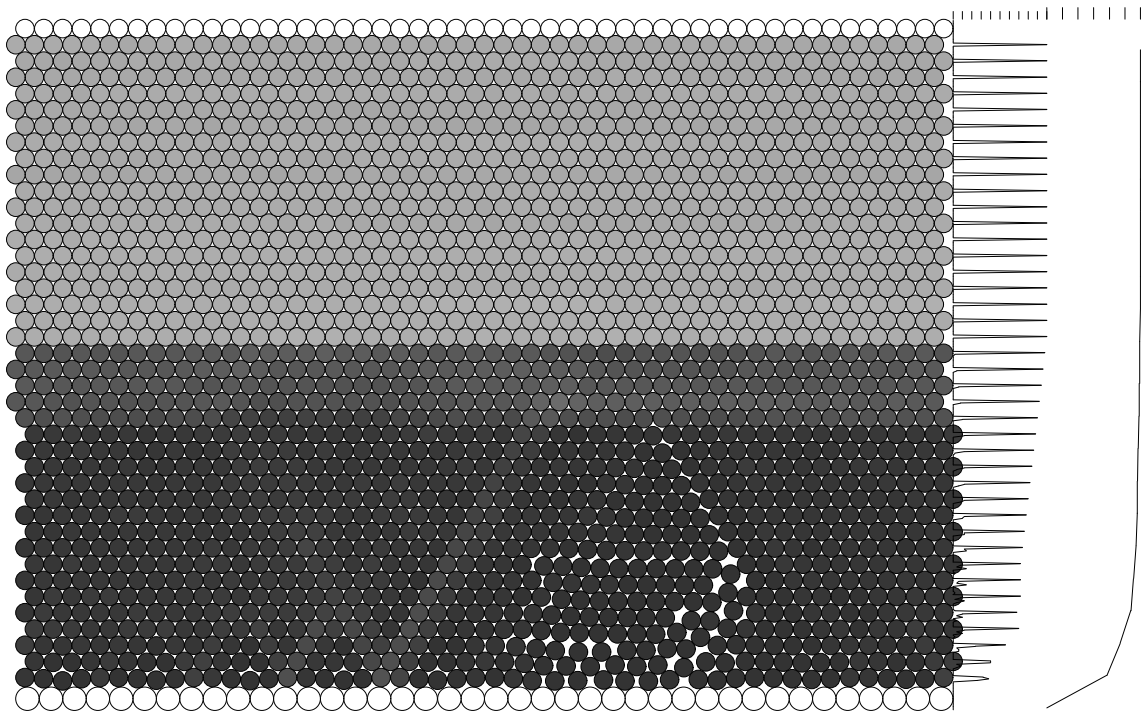}
\caption{
Final jammed configuration for case H, 
exhibiting geometric frustration over at least the bottom half 
of this 1990-mobile-disk system
}
\label{1caseh}
\end{figure}

\begin{figure}
\includegraphics*[width=6.2in]{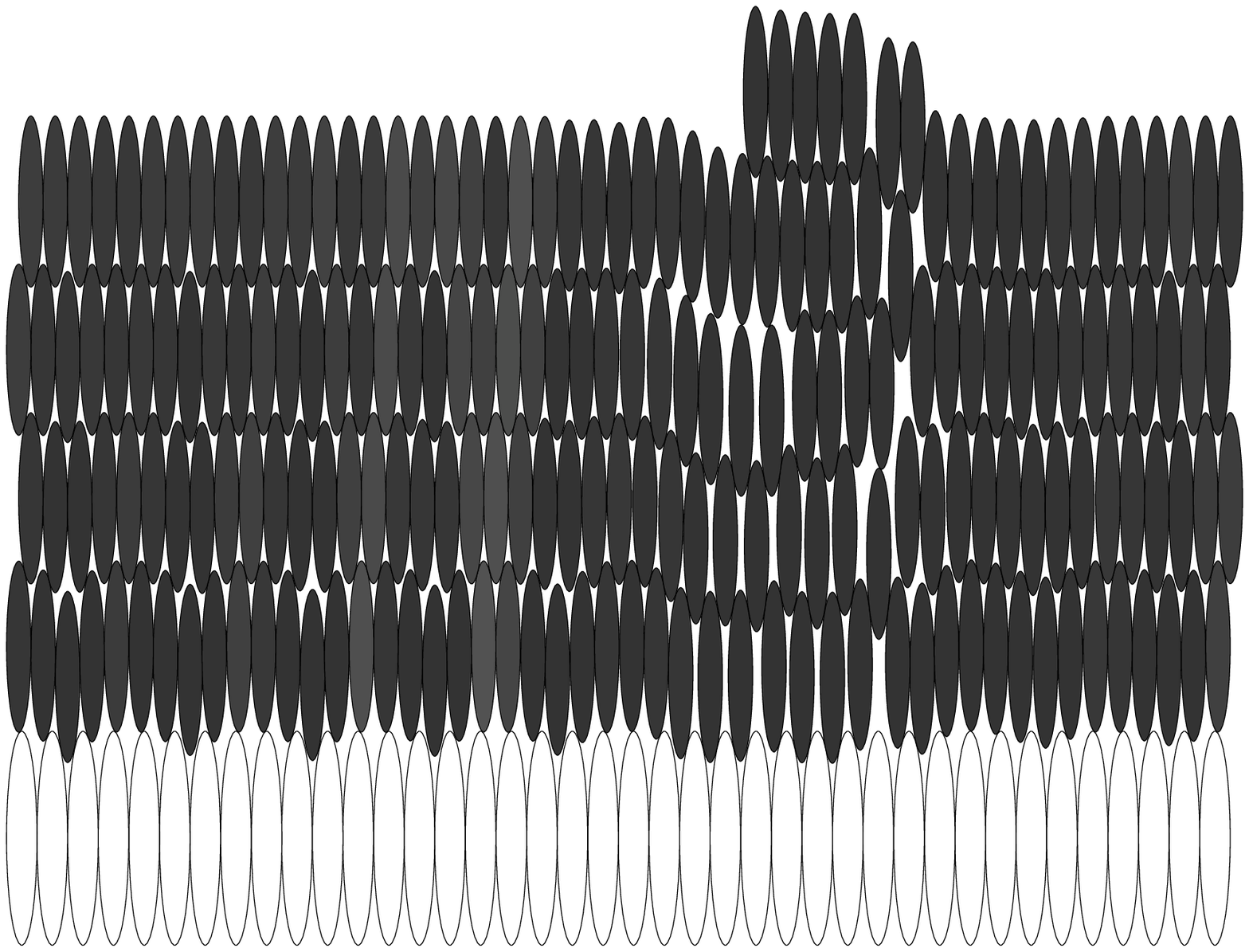}
\caption{
Vertically stretched view of the bottom layers 
in the jammed configuration for case H, 
shown in the previous Figure ~\ref{1caseh}
}
\label{7caseh}
\end{figure}

\begin{figure}
\includegraphics*[width=6.2in]{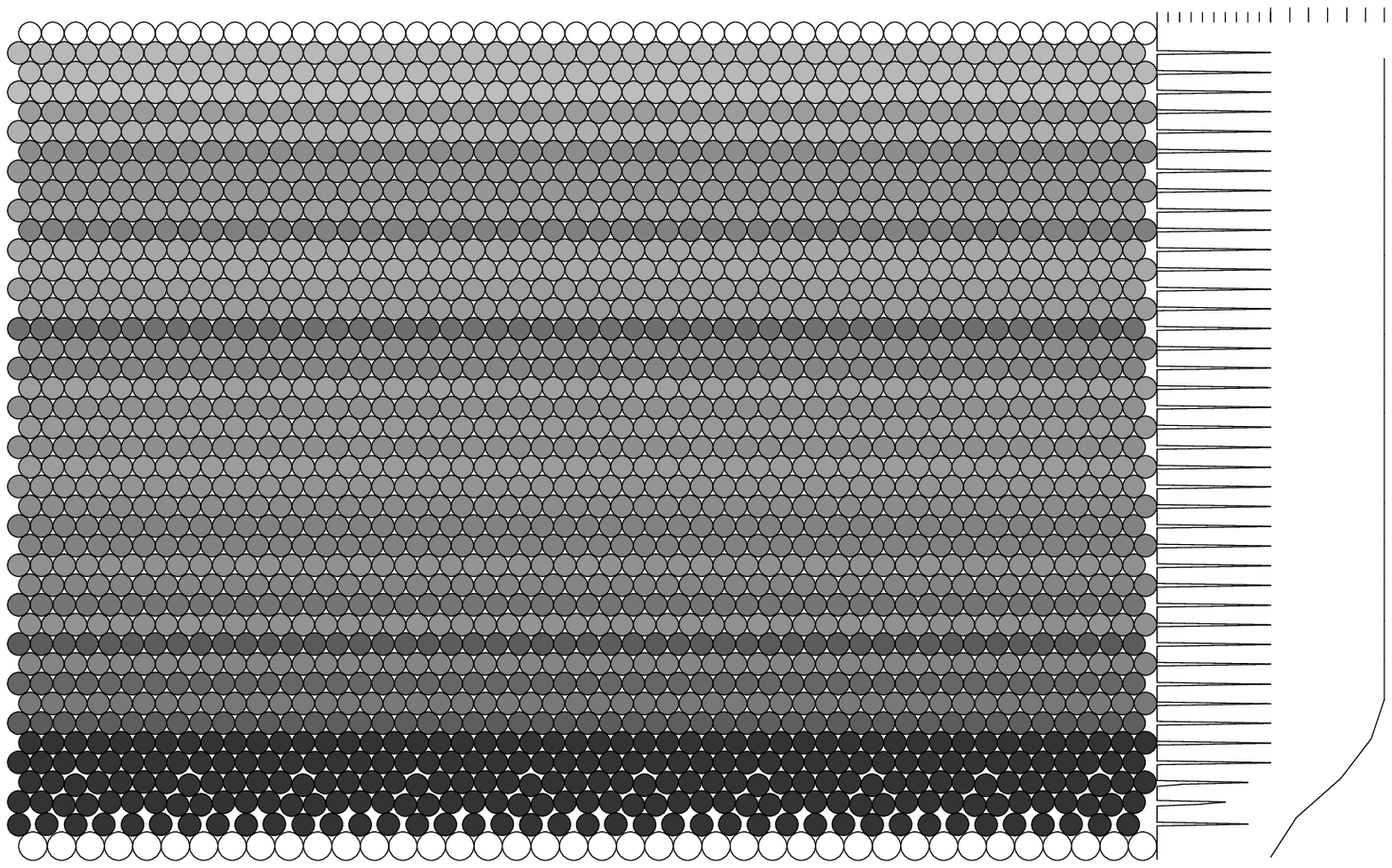}
\caption{
Densest final configuration for the case of 1990 movable disks.  
This pattern has a horizontal period of 4 length units, 
and was created by applying 
a restricted configurational search as described in the text
}
\label{best1990}
\end{figure}
In contrast to the 2000-disk cases, 
none of the final configurations for 
1990-disk cases F to J looks 
even close to periodic in the horizontal direction.  
However, 
this does not preclude a search for horizontal 
period-4 solutions.  
All of the conditions 
for such a restricted configuration space search 
would be applied as before, 
except that 199 mobile disks would be involved instead of 200.  
One possible outcome is that 
the final configuration 
would be the same as for 200 mobile disks, 
but with a single monovacancy 
(missing disk) 
located somewhere within the structure, 
and with a corresponding reduction in density.

Perhaps surprisingly, 
a modest search of the type just described in fact discovered 
a period-4 packing 
applicable to any multiple of 199 disks, 
in particular to 1990.  
Its density is clearly higher than those of 
cases F-J reported in 
Table~\ref{tab1}, 
specifically  
$\xi(N=200)~=~0.900088$, and  
$\hat{\xi}(N=200)~=~0.899897$, 
and it is also higher 
than that of the configurations 
of 200 disks with a monovacancy.
However, 
the found 199-particle density is noticeably lower 
than the corresponding 200-particle highest density mentioned above.  
Figure ~\ref{best1990} 
displays the configuration, 
periodically replicated 10-fold 
to produce a 1990-particle example. 
In contrast to the previous 200-disk periodic structure, 
the bottom layer of movable particles consists in 
each period of only four individuals 
rather than five, 
each buried in a pocket provided by the fixed larger disks.  
This accounts for the missing particle, 
rather than by monovacancy formation.  
The subsequent 
layers proceeding vertically contain five disks, 
but in a wavy pattern to contact the four below.  
In a fashion similar 
to that of the 200-particle periodic case, 
the amplitude of waviness diminishes with altitude, 
and is virtually gone by the fifth layer.  
And like the 200-particle periodic case, 
the suspected 
optimum packing for 199 particles 
apparently also possesses mirror symmetry 
with respect to vertical axes 
separated from one another by one-half period 
(i.e. 2 units in the $x$ direction).

Based on this last finding, 
we speculate that the densest packing 
in the case of 1990 mobile particles is 
periodic in the horizontal direction with period 4, 
and indeed is the structure illustrated in 
Figure ~\ref{best1990}.  
It should be noted in passing 
that this structure was obtained only after 
15 unsuccessful attempts 
to create a suitably high density result.  
Each of the 15 preceding unsuccessful attempts 
produced a configuration 
identifiable as a 
monovacancy-containing modification of 
a 200-particle structure, as described above. 
This is consistent with the earlier remark 
that apparently a substantially greater diversity of jammed 
structures exist, 
most of relatively low density, 
in the 1990-particle system in comparison to the 2000-particle system.     

\section{Numerical Results, Spheres}
\hspace*{\parindent}
The behavior of rigid disks 
in epitaxial jamming circumstances, 
described in the preceding 
Section 3, 
is already complex and challenging.  
The analogous situation for spheres can be expected to be even more so.  
In part this arises from the fact that, 
even disregarding boundary effects, 
spheres possess a larger 
relative number and greater geometric diversity of packing types 
\cite{LS, LSP, DTSC}.  
But it also stems from the
greater number of periodic substrate structures 
that could be presented to the movable sphere set.  
The 
choices described in 
Section 2 
above to limit the sphere simulations 
to square arrays for bottom and top rigid layers, 
and to use slow compression, 
were motivated by the need to keep the analysis as tightly 
focused as possible. 
 
The final densities for the two sphere cases 
considered are the following: 
\begin{equation}
\label{sphden}
\begin{array}{rl}
{\xi(N=18000)~=~0.679751},
& 
{\hat{\xi}(N=18000)~=~0.676383};          
\\
{\xi(N=17541)~=~0.680724},
&
{\hat{\xi}(N=17541)~=~0.677318}.
\end{array}
\end{equation}
In spite of the use of slow compression, 
the final jammed configurations 
for both of these cases exhibit 
substantial disorder pervading 
nearly the entire interval 
between bottom and top rigid layers.  
Figures ~\ref{yz18000} and ~\ref{yz17541} 
present typical planar $y,z$ slices for the two cases.  
These figures include collision-rate shading as 
in the prior disk cases.  
They also present $z$-direction density histograms, 
and binned contact and near-contact numbers 
for all movable spheres, 
at the right edges.
The two sets of ticks at the top edges of these two graphs
allow one to deduce the magnitude of the values in the graphs.
The arrangement of the ticks is analogous to that
in Figure ~\ref{1cased} and it was described in Section 3 above.
In particular, the spacing between consecutive ticks
for the $z$-direction density histogram is 10\% of the
maximum number of particle centers that may occur at a given $z$
under the planar square arrangement,
i.e., 10\% of $30~ \times~ 30~ =~ 900$.
Thus, the second tick corresponds to 10\%, i.e., to 90 particles in a bin,
the third to 20\%, i.e., to 180 particles, and so on.
\begin{figure}
\includegraphics*[width=6.4in]{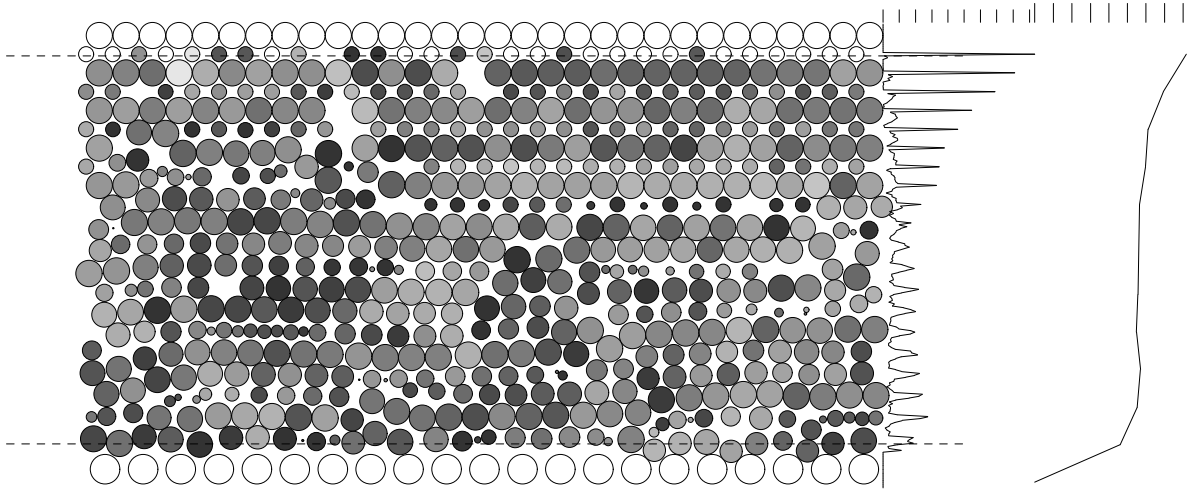}
\caption{
$y,z$ planar slice (at $x~=~0.75300~L_x$) 
through the final jammed configuration 
for the 18000-movable-sphere case.  
The horizontal dashed lines near bottom and top 
locate slices used for Figures 
~\ref{xybot18000} 
and 
~\ref{xytop18000} 
}
\label{yz18000}
\end{figure}

\begin{figure}
\includegraphics*[width=6.4in]{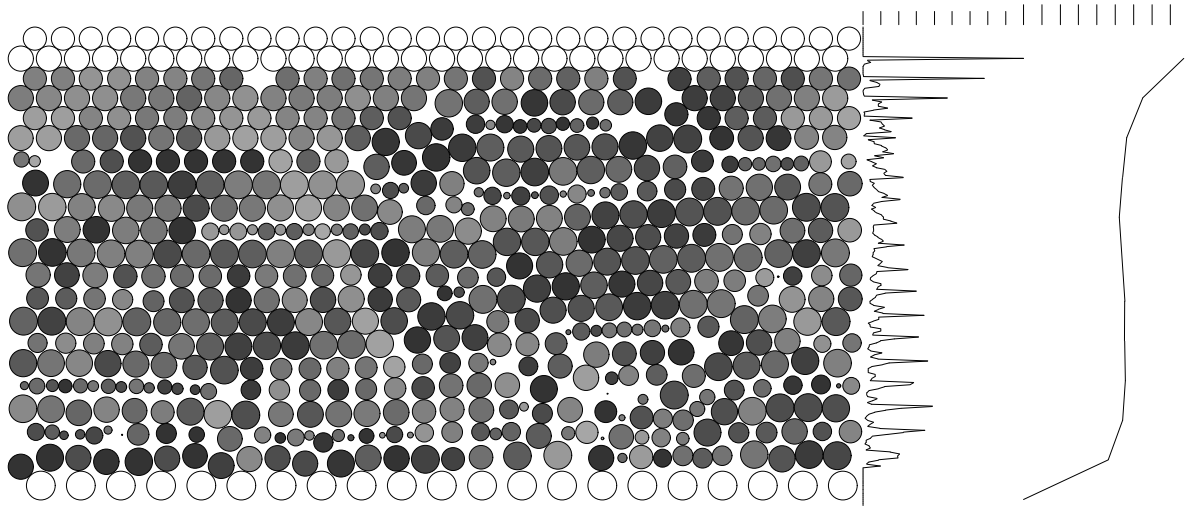}
\caption{
$y,z$ planar slice (at $x~=~0.32600~L_x$) 
through the final jammed configuration 
for the 17541-movable-particle case
}
\label{yz17541}
\end{figure}

\begin{figure}
\includegraphics*[width=5.3in]{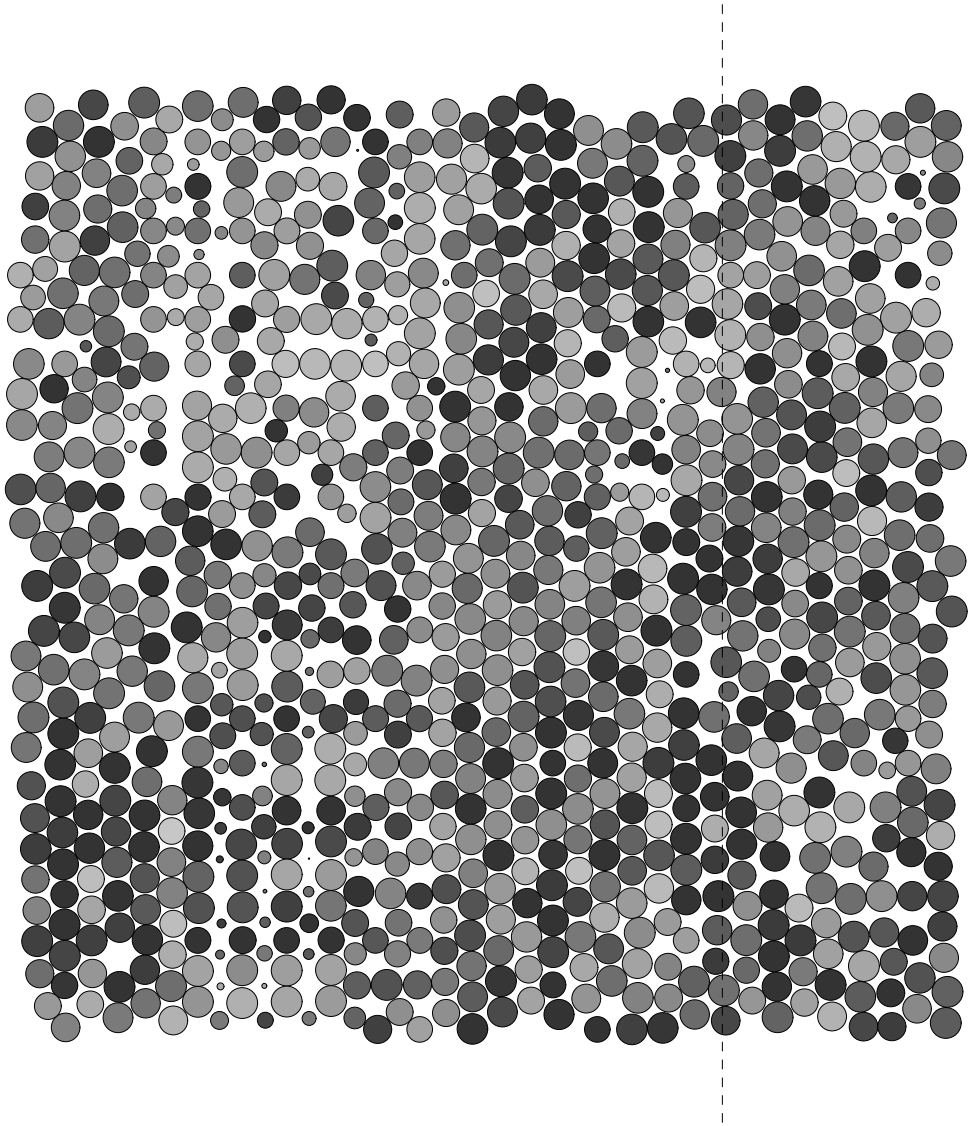}
\caption{
$x,y$ planar slice through the final jammed configuration 
for the 18000-movable-particle case.  
This view represents a portion of the sphere arrangement 
near the bottom substrate layer of unit-diameter fixed spheres, 
at 1.027 distance units above the plane 
containing the centers of those fixed spheres.  
The dashed line indicates the position of the $y,z$ plane 
used for the slice shown in Figure ~\ref{yz18000}
}
\label{xybot18000}
\end{figure}

\begin{figure}
\includegraphics*[width=5.3in]{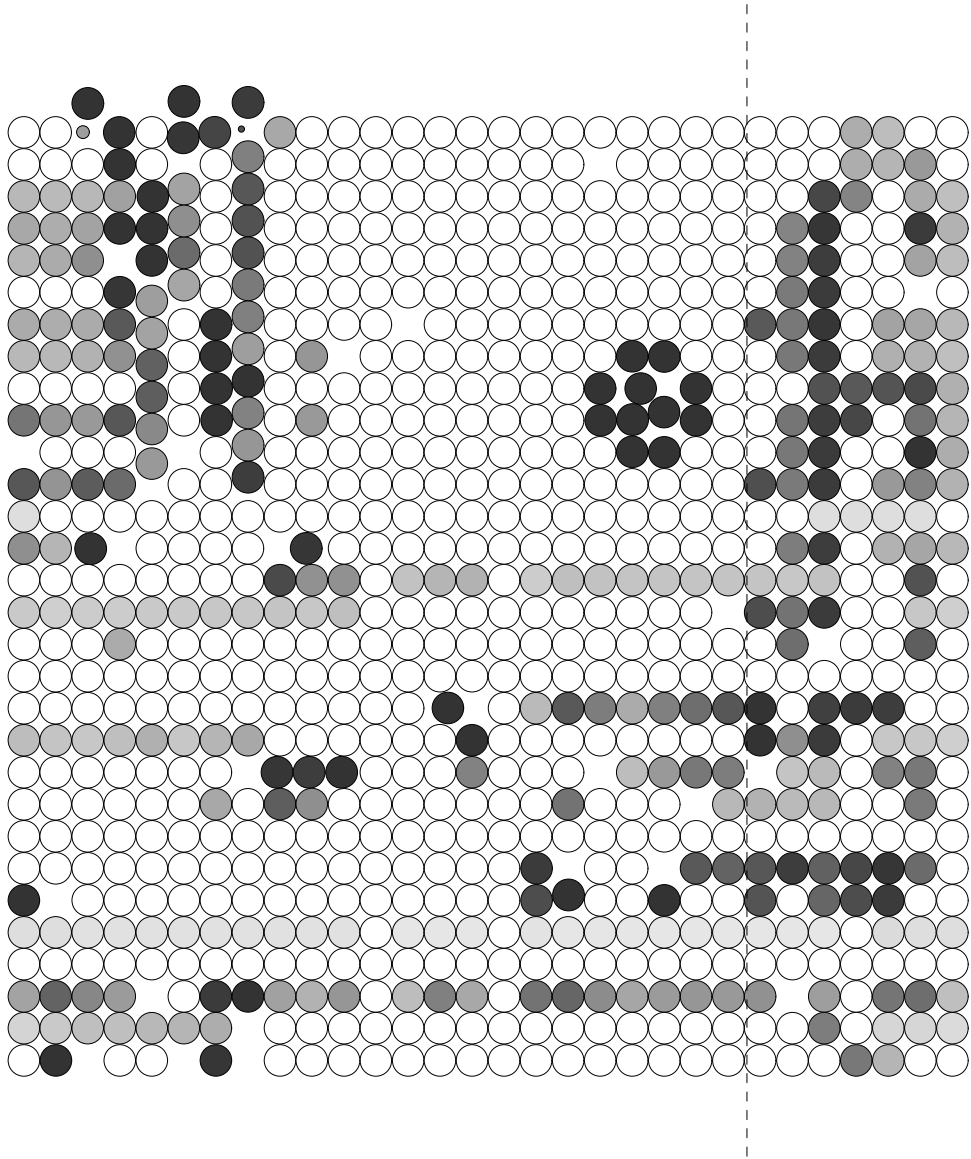}
\caption{
$x,y$ planar slice through the final jammed configuration 
for the 18000-movable-particle case.  
This view represents a portion of the sphere arrangement 
near the rigid top layer of spheres that have the same diameter 0.7 
as the movable spheres.  
The slice resides 0.772 distance units below the plane 
containing the centers of spheres comprising that rigid top layer.  
The dashed line indicates the position of the $y,z$ plane 
used for the slice shown in Figure ~\ref{yz18000}
}
\label{xytop18000}
\end{figure}
  
Figures ~\ref{xybot18000} and ~\ref{xytop18000}
illustrate further aspects of the 18000-sphere case 
by showing a pair of horizontal slices 
(i.e. at fixed altitudes $z$).  
The former involves a slice near the bottom substrate layer of rigidly fixed 
incommensurate spheres, 
while the latter shows the result 
for a slice near the top layer of rigidly fixed 
small (commensurate) spheres.  
The positions of these slices are indicated by dashed lines in 
Figure ~\ref{yz18000}.  
These views confirm an attribute suggested by the prior 
Figure ~\ref{yz18000}.  
that the extent of configurational 
disorder diminishes somewhat from bottom to top.  
This is reasonably clear from 
Figures ~\ref{xybot18000} and ~\ref{xytop18000}
not only by the relative extents of square packing order, 
but also by the diagnostic shading of the particle 
images, 
which we saw in the disk case correlates with disorder.  
Although we do not show them, 
similar fixed-$z$ slices for the 17541-movable-sphere case 
present the same qualitative scenario.

In view of the observations in the preceding 
Section 3 
that disk simulations tend to generate periodic 
or near-periodic patterns 
in contact with the incommensurate bottom layer, 
it is natural to look for a 
similar behavior with spheres.  
Given the radius ratio 0.7, 
and the square (001) geometry of the fixed 
bottom layer, 
the sphere analog would be 
a horizontally doubly-periodic pattern with 
size $7 \times 7$ units 
(i.e. $10 \times 10$ diameters 
of the smaller movable spheres).  
However, examination of the 
generated configurations by means of graphical presentations 
such as those in 
Figures ~\ref{xybot18000} and ~\ref{xytop18000}
reveals 
no obvious tendency toward such doubly-periodic patterns.  
This may be due to the significantly larger 
number of movable particles that would have to be involved, 
as well as due to the disruptive influence of 
spontaneously formed disordered structures within the bulk 
of the compressed layer.  
Failure to observe 
horizontal double periodicity in our limited search 
for the 18000 and 17541-sphere systems 
thus should not be interpreted as evidence against 
the existence of such a possibility, 
but merely that it is improbable.

Motivated by analogous success 
in finding high-density periodic 
patterns for the disk systems, 
we have 
performed several additional sphere runs 
where the system geometry was periodically restricted in each of 
the $x$ and $y$ directions by a factor of 3.  
Thus, 
for a relevant doubly-periodic subset of the 18000 sphere 
system we considered configurations of only 2000 mobile spheres, 
while for the 17541 sphere system we 
examined configurations of 1949 mobile spheres.  
In both of these 
searches   
$L_x=L_y$
was 
restricted to 7 units, 
so that the rigid top layer involved
$10 \times 10~=~100$
small spheres, 
and the rigid bottom layer consisted of
$7 \times 7~=~49$
large spheres.  
As was the case for disks, 
the highest densities obtained in these restricted 
circumstances exceeded the corresponding values 
for the unrestricted systems.  
This provides at least 
weak circumstantial evidence 
that the highest possible densities require double periodicity.  
However, 
we hasten to point out 
that no easily described regular patterns emerged.  
On the contrary, 
we observed that 
the final configurations for these smaller systems 
contained many structural irregularities, 
and to that 
extent were not directly analogous to the disk cases.  
Finding structurally ``perfect'' doubly periodic 
patterns by this approach, 
if they exist, 
presumably would require a substantially larger number of search 
attempts, 
and would exceed 
the scope and available resources of the present investigation. 

\section{Conclusions and Discussion}
\hspace*{\parindent}
Computer simulation 
has been applied to the study of selected aspects of epitaxial frustration 
phenomena for systems of hard disks and hard spheres 
forced into jamming contact with rigid ordered 
layers that are composed of larger disks and spheres, respectively.  
In order to keep this inquiry within 
reasonable limits, 
single rational size ratios were selected 
(0.8 for disks, 0.7 for spheres).  
Furthermore, 
only structurally perfect arrays served as substrates: 
a linear array for disks, and a (001) layer
from a face-centered cubic close packed array for spheres. 
Jamming was produced 
by slow vertical compression onto the incommensurate substrate, 
by means of constant slow-speed approach 
of the incommensurate lower and commensurate upper
boundary layers. 
Periodic boundary conditions applied in directions 
parallel to the boundary layers.

Two sizes of disk systems 
(2000 and 1990 movable particles) were simulated. 
Both produced final structures 
that displayed a diversity of patterns 
differing in the amount and location of packing disorder. 
Nevertheless, 
results indicated that the most ``efficient'' arrangements of disks
(i.e. those exhibiting the highest final density)
possessed a pattern periodic in the direction parallel to the substrate, 
with most of the geometric disturbance 
of the natural triangular lattice packing of disks 
localized close to the incommensurate substrate.

The sphere systems investigated, 
containing 18000 and 17541 movable particles, 
appeared to settle into more disordered jammed arrangements 
in comparison with the disk systems. 
In part this could be attributed to the known bulk properties 
of the disks and spheres, 
namely that far more geometrically stable amorphous packings 
exist for the latter than for the former 
\cite{LS, LSP, DTSC}.  
A search for high-density doubly-periodic sphere packings 
experienced limited success; 
while relatively high densities in fact appeared, 
the packings displayed considerable disorder beyond what 
would be expected in the immediate vicinity 
of the incommensurate substrate layer. 
Consequently, 
it remains an attractive prospect for a future investigation 
to attempt to identify ideal 
doubly-periodic packings with minimum disorder 
and maximum attainable density. 
Success in such an attempt 
may require use of special initial conditions 
that strongly diminish the occurrence of sphere packing disorder 
away from the substrate layer.

The substrate geometries 
employed in the present study 
constitute a very small subset of all 
possibilities that might be considered.  
These possibilities include substrates that are crystalline alloy 
structures with two or more distinct particle sizes, 
stepped substrate surfaces, 
and substrates containing 
various point and extended defects.  
Even amorphous substrates 
presenting a substantially planar surface 
to the movable particles deserve attention.
All of these extensions present varying degrees
of reduced order to the movable particles,
and it will be illuminating to determine
the characteristic types and the extent
of extra packing disorder in the deposited layers
that they indeividually produce.

With respect to the more modest goal 
of extending the present work to a wider range of radius ratios, 
several points deserve at least brief discussion. 
In principle, 
the full range of size ratios 
from very small to very large could be examined for disks, 
going well beyond the single ratio 0.8 considered in this paper. 
Of course, 
imposition of periodic boundary conditions 
in the direction parallel to the fixed substrate 
effectively constrains those ratios to be simple rational numbers. 
Epitaxial frustration will inevitably arise, 
except when the movable disks have diameters 
equal to an integer multiple (1, 2, 3, ....) 
of those comprising the fixed substrate.  
When these integer multiples are present,
larger movable disks can form a perfect contacting row by 
settling respectively into every 
first, second, third, .... pocket 
presented by the substrate, 
and this can serve as the bottom row 
of a perfect triangular crystal arrangement 
for the remainder of the movable disks. 
 
The range of available size ratios 
for the three-dimensional case of spheres is a bit more restricted. 
The movable spheres cannot be so small 
that they could slip between the fixed spheres that form the 
substrate layer.  
In the case of a (001) face-centered cubic structure 
as used in the simulations reported above, 
this requires that the movable spheres be no smaller 
than   
$\sqrt{2}~-~1~=~0.41421...$
times 
the substrate sphere size.  
For a substrate possessing the close-packed (111) surface structure, 
the corresponding ratio would be substantially smaller 
but still positive, 
specifically
$1/\sqrt{3}~-~1/2~=~0.07735...$. 
As in the disk case, 
a discrete set of movable-to-substrate size ratios 
greater than unity can avoid epitaxial 
frustration by allowing perfect fit of 
a first added layer into a subset of pockets provided by the substrate, 
while serving as the beginning of a structurally perfect crystal 
of those larger spheres.

Finally, 
it is appropriate to mention 
that careful theoretical examination of packing geometry for bulk 
disk and sphere systems has revealed 
the importance of distinguishing various definitions of ``jamming'' 
\cite{TS, TDS, DTSC, SST}. 
Considering the method used in our simulations, 
the natural presumption would be that the final disk or 
sphere states obtained are at least close to the 
``collectively jammed'' state, 
that is (aside from inclusion of occasional ``rattlers''),
the initially movable particles have reached 
a state such that no subset of them can be simultaneously 
shifted while respecting the non-overlap constraints.  
The more stringent ``strictly jammed'' category 
however may not apply; 
that would require testing for 
system instability by deforming 
the rectangular system boundaries 
to parallelogram or parallelopiped shape 
by application of shear parallel to the substrate. 
As is the case for the other directions of extended investigation, 
this aspect of the general problem will 
have to be reserved for the future.

\end{document}